%% file: splatbus.tex
\documentclass[sigconf]{acmart}

\usepackage{multirow} %

\usepackage[ruled]{algorithm2e} %
\usepackage{subcaption}
\newcommand{\name}{SplatBus\xspace}

\copyrightyear{2026}
\acmYear{2026}
\setcopyright{cc}
\setcctype{by}
\acmConference[SA Technical Communications '26]
  {SIGGRAPH Asia 2026 Technical Communications}
  {December 01--04, 2026}
  {Kuala Lumpur, Malaysia}
\acmBooktitle{SIGGRAPH Asia 2026 Technical Communications
  (SA Technical Communications '26), December 01--04, 2026,
  Kuala Lumpur, Malaysia}
\acmDOI{10.1145/3829339.3847851}
\acmISBN{979-8-4007-2841-9/2026/12}

\begin{document}
\title{\name: A Gaussian Splatting Viewer Framework via GPU Interprocess Communication}

\author{Yinghan Xu}
\orcid{0009-0008-7682-8070}
\affiliation{%
 \institution{Trinity College Dublin}
 \city{Dublin}
 \country{Ireland}}
\email{yixu@tcd.ie}
\author{Théo Morales}
\orcid{0000-0002-2275-0895}
\affiliation{%
 \institution{Trinity College Dublin}
 \city{Dublin}
 \country{Ireland}}
\email{tmorales@tcd.ie}
\author{John Dingliana}
\orcid{0000-0001-7636-6402}
\affiliation{%
 \institution{Trinity College Dublin}
 \city{Dublin}
 \country{Ireland}}
\email{john.dingliana@tcd.ie}

\input{sec/0_abstract}



\maketitle

\input{samplebody-journals}

\end{document}

%% file: sec/0_abstract.tex
\begin{abstract}
Radiance field-based rendering methods have attracted significant interest from
the computer vision and computer graphics communities. They enable high-fidelity
reconstruction with complex real-world lighting effects, but at the cost of high
rendering time. 3D Gaussian Splatting solves this issue with a
rasterization-based approach for real-time rendering, enabling applications such
as autonomous driving, robotics, virtual reality, and extended reality. However,
due to the lack of format and rendering standardization, the ecosystem is
fragmented into specific viewers. This makes it difficult to integrate these
methods into traditional mesh-based rendering pipelines, which is a common use
case for interactive applications. To address this limitation, we propose a
framework that uses Nvidia’s interprocess communication (IPC) APIs to provide an
implementation-agnostic rendering server, enabling the results to be viewed in
external clients such as Unity, Blender, and OpenGL viewers.
Code is available at \url{https://github.com/RockyXu66/splatbus}.
\end{abstract}

%% file: samplebody-journals.tex
\input{sec/0_figures}

\input{sec/1_intro}

\input{sec/2_background}

\input{sec/3_method}

\input{sec/4_evaluation}

\input{sec/5_conclusion}

\begin{acks}
This work was conducted with the financial support of the Research Ireland
Centre for Research Training in Digitally-Enhanced Reality (d-real) under
Grant No. 18/CRT/6224.
\end{acks}

\bibliographystyle{ACM-Reference-Format}
\bibliography{sample-bibliography}

%% file: sec/0_figures.tex
\begin{figure}[tb]
    \centering
    \includegraphics[width=\columnwidth]{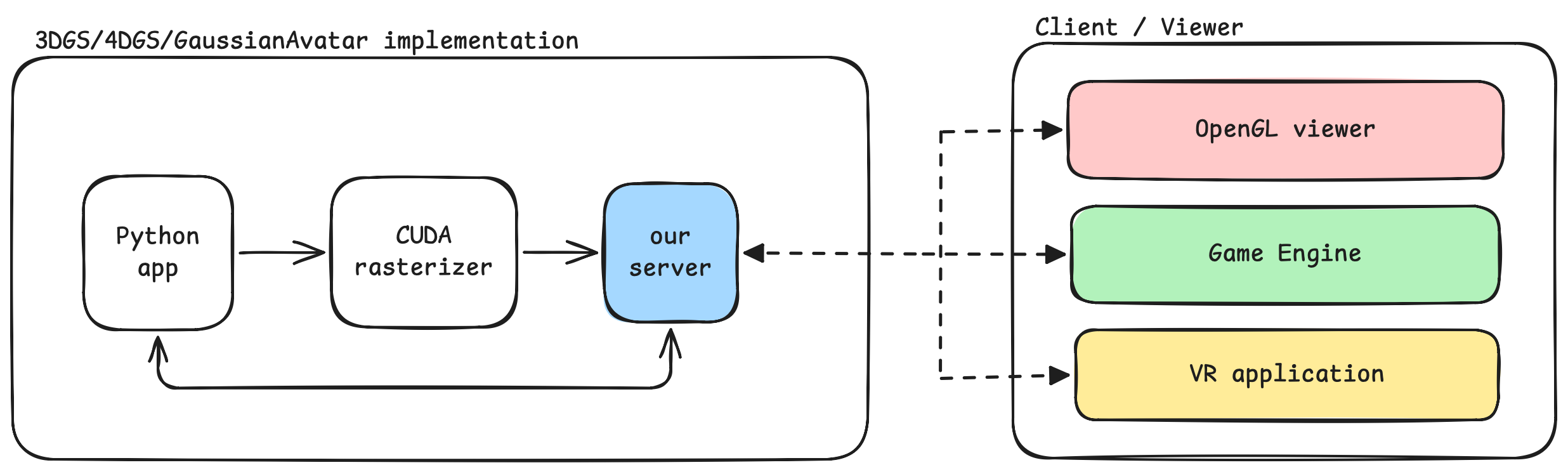}
    \caption{Overview of \name. A Python rendering server exposes rendered color
    and depth buffers through CUDA IPC, while external clients exchange camera
    and scene controls over sockets.}
    \label{fig:system}
\end{figure}

\begin{figure*}[t]
    \centering
    \raisebox{-1em}{
    \begin{subfigure}{0.32\textwidth}
        \centering
        \includegraphics[width=\textwidth]{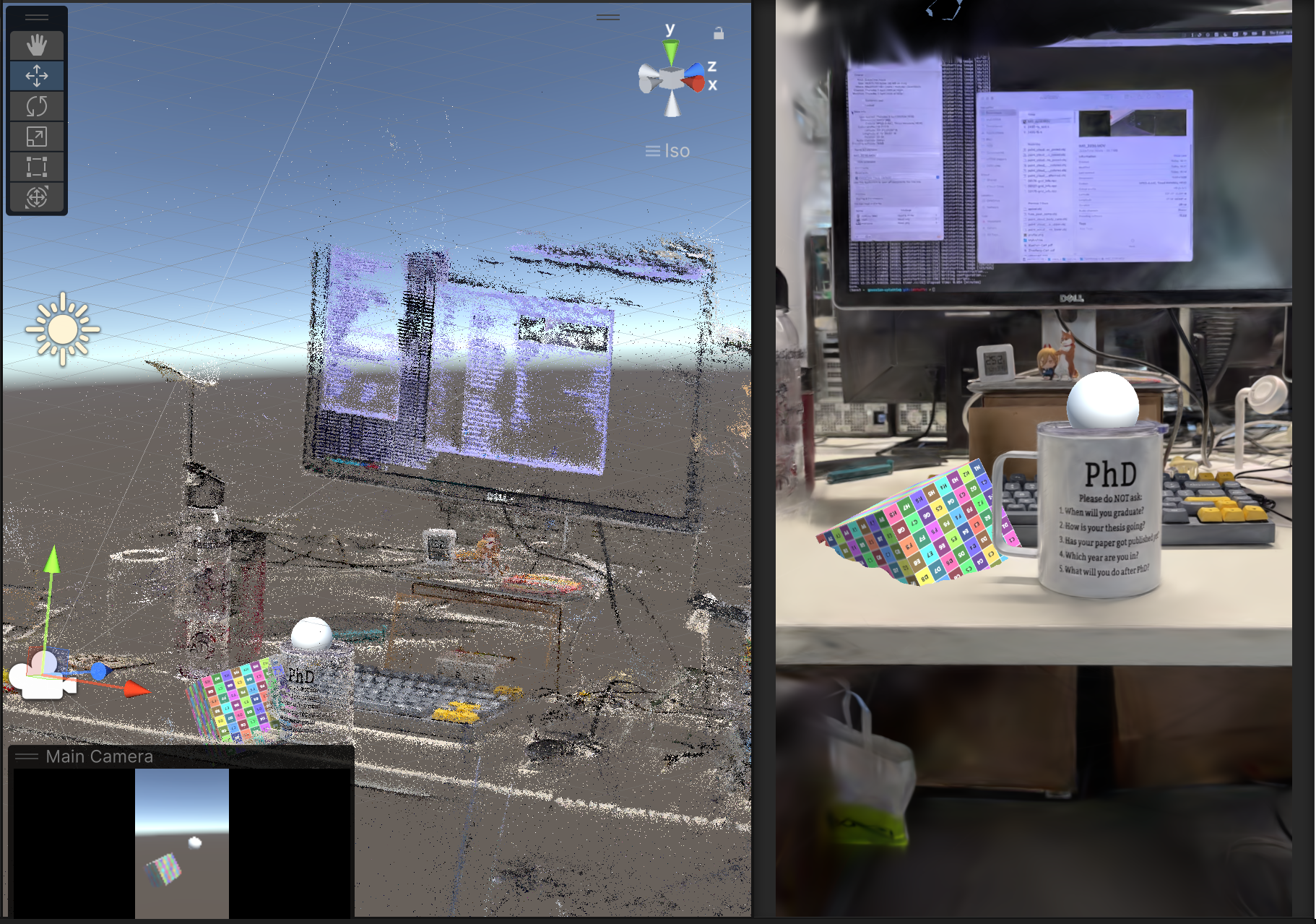}
        \caption{Unity plugin -- Left: 3DGS scene point cloud in the Scene view.
        Right: 3DGS rasterizer output composited on the Unity scene in the Game view.}
    \label{fig:unity-screenshot-3dgs}
    \end{subfigure}
    }
    \hfill
    \begin{subfigure}{0.38\textwidth}
        \centering
        \includegraphics[width=\textwidth]{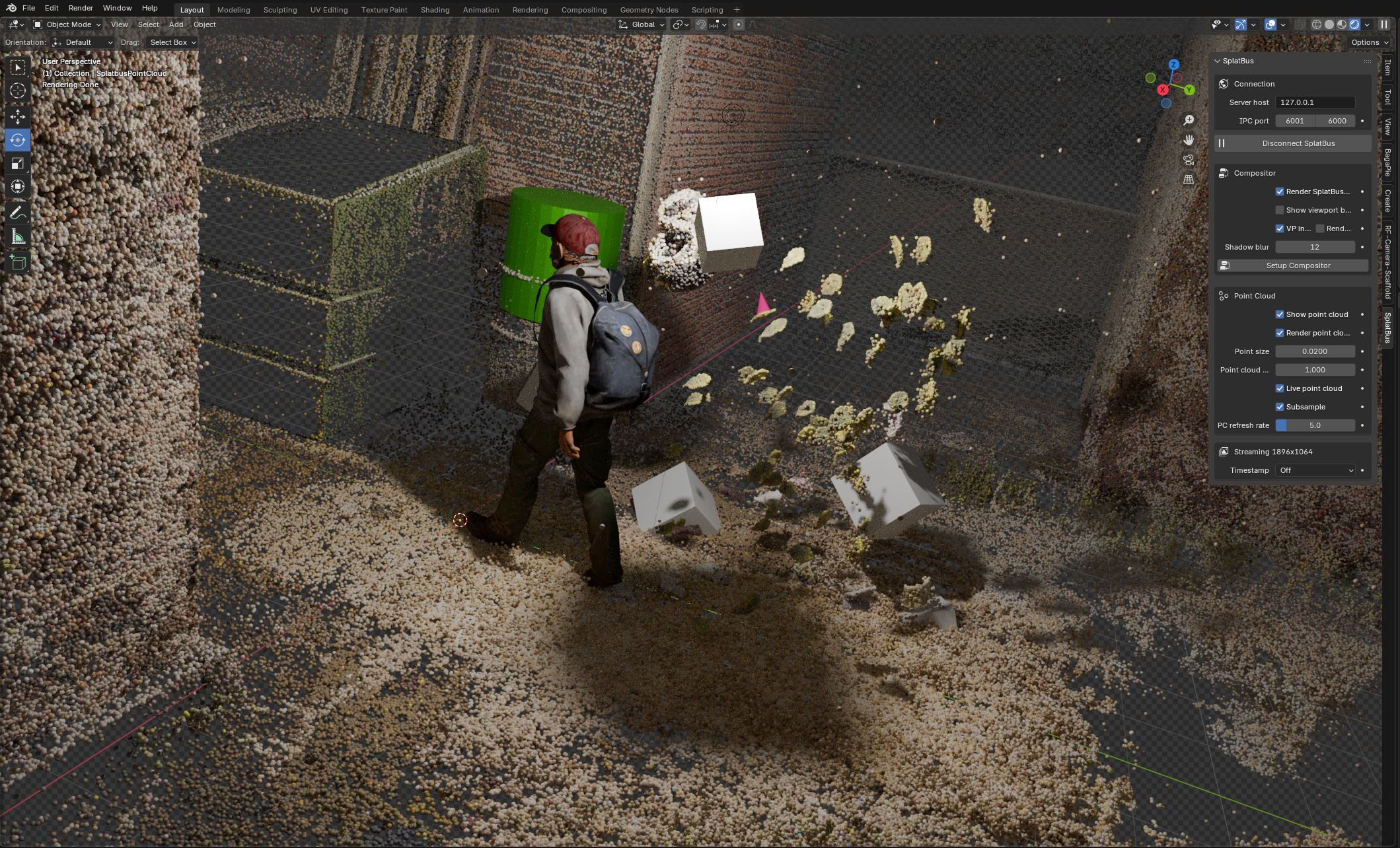}
        \caption{Blender plugin -- visualization of a 3DGS coloured pointcloud casting shadows
        onto a mesh character in the viewport.}
        \label{fig:blender-screenshot}
    \end{subfigure}
    \hfill
    \begin{subfigure}{0.24\textwidth}
        \centering
        \includegraphics[width=0.5\textwidth]{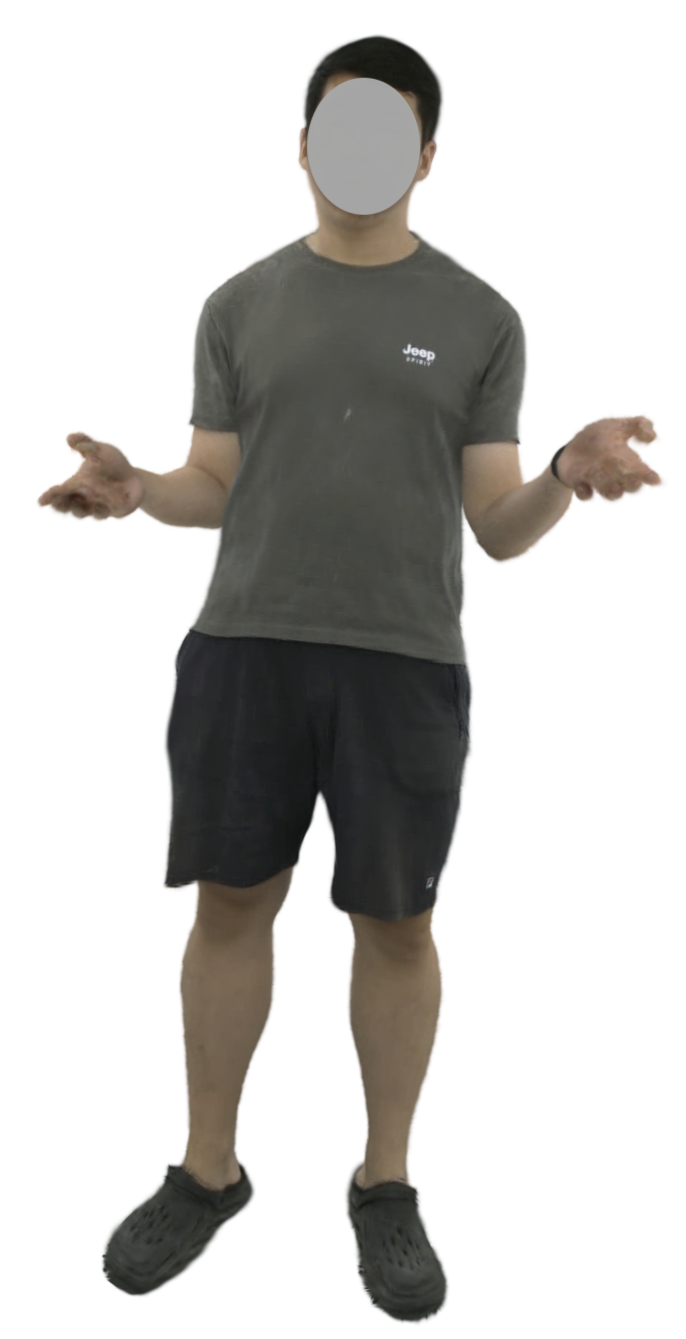}
        \caption{OpenGL viewer -- visualization of animated Gaussian Avatar renders.}
    \label{fig:opengl-screenshot-avatar}
    \end{subfigure}
    \caption{Screenshots of SplatBus clients showing plug-and-play integration
    with various 3DGS-based methods and compositing with mesh rendering pipelines.}
    \label{fig:clients}
\end{figure*}

%% file: sec/1_intro.tex
\section{Introduction}
\label{sec:intro}

Neural Radiance Fields (NeRFs) popularized high-quality novel view synthesis
from multi-view images~\cite{10.1145/3503250},
although they are known for their costly inference. While this is addressed
with specialized acceleration structures and heuristic-based implementations~\cite{10.1145/3528223.3530127}, the NeRF framework remains too
costly for real-time renders.
3D Gaussian Splatting
(3DGS)~\cite{10.1145/3592433} instead represents scenes as rasterizable Gaussian
primitives, providing an attractive balance of quality and speed. Yet, its
viewing ecosystem remains fragmented: the original SIBR viewer~\cite{sibr_framework}
is a stand-alone C++ application coupled to a CUDA rasterizer, while newer 3DGS
methods often modify the rasterizer and must adapt the viewer in parallel. This
coupling creates substantial engineering overhead, as seen in dynamic
radiance fields, avatar methods, or pipelines with neural components before
rasterization.

Game-engine and web integrations enable useful deployment workflows, including
hybrid rendering with meshes and UI~\cite{playcanvas_supersplat,pranckevicius2023unitygs},
but they usually reimplement a particular 3DGS rasterizer or convert splats into
engine-specific representations. As a result, they can lag behind research-oriented
rasterizer variants, struggle with non-standard Gaussian-based methods (i.e. 2D
Gaussian Splatting~\cite{10.1145/3641519.3657428}), or require repeated
integration effort across viewers.

We propose \name, a client--server viewer framework that decouples Gaussian-based
rendering from visualization. The research renderer runs unchanged as a
Python server, while clients such as Unity, Blender, or OpenGL viewers
receive rendered color and depth buffers through NVIDIA CUDA inter-process
communication (IPC). By sharing GPU buffers and synchronization events directly,
\name avoids CPU readback during frame handoff, preserves the renderer's exact output, and exposes depth for hybrid composition with mesh-based content. This design turns viewer
integration into a lightweight API hook rather than a renderer rewrite, making
new Gaussian-based research code usable in interactive external environments.

%% file: sec/2_background.tex
\section{Related work}
\label{sec:relatedwork}

Existing Gaussian Splatting viewers are typically tied to a specific renderer or
runtime. The original 3DGS release provides an interactive SIBR-based
viewer~\cite{sibr_framework}; web viewers and editors such as
\emph{GaussianSplats3D}~\cite{kellogg_gaussiansplats3d} and
\emph{SuperSplat}~\cite{playcanvas_supersplat} make static 3DGS scenes easier to
view, edit, and deploy in browsers; and NerfStudio~\cite{nerfstudio} supports a
client--server workflow for radiance-field visualization. Dynamic-scene systems
such as EasyVolcap~\cite{xu2023easyvolcap} and real-time 4DGS
viewer~\cite{realtime_4dgs} support dynamic scenes, but adapting
these viewers to modified rasterizers, neural network components, or
new Gaussian variants often requires rewriting the viewer itself.

Game-engine integrations provide richer interaction, Virtual Reality (VR), and
hybrid rendering with meshes and UI.
UnityGaussianSplatting~\cite{pranckevicius2023unitygs} and
gsplat-unity~\cite{gsplatunity} reimplement 3DGS in Unity compute shaders, while
XScene-UE\-Plugin~\cite{xverse2024xscene} brings splat rendering and editing to
UE5. However, these integrations are usually coupled to one engine and one
rasterization path. In contrast, our approach keeps the original research
renderer as an independent Python server and uses GPU IPC to share its rendered
color and depth buffers with multiple clients, preserving renderer fidelity
while enabling external visualization and hybrid scene composition.

%% file: sec/3_method.tex
\section{Method}
\label{sec:method}

\name decouples Gaussian-based rendering from interactive visualization by
turning the original research renderer into a lightweight Python server
(Fig.~\ref{fig:system}). Instead of reimplementing a rasterizer inside each
viewer, the server keeps the user's rendering loop and publishes its
rendered color and depth images to external clients. Unity, Blender, and OpenGL
clients share the same protocol, so new viewers can be added without changing
the Gaussian rasterizer or its related research code snippet.

\subsection{Python rendering server}

The server package is installed on top of an existing Gaussian rasterization
environment and exposes two GPU-resident frame buffers, with color in RGBA8 and
depth in R32F. Both buffers are allocated as contiguous CUDA memory and updated
directly on the GPU after each render. When the underlying method outputs
inverse depth, the server converts it to linear depth so clients can perform
standard depth-aware composition. For cross-process sharing, the server creates
CUDA IPC memory handles for the color and depth buffers plus an IPC-capable CUDA
event. After writing a frame, it records the event, allowing clients to wait for
completion before reading the shared buffers.

A pair of lightweight TCP channels handles non-image data. During initialization,
the server sends buffer metadata and CUDA IPC handles to the client. During
interaction, the client streams camera and object transforms, which the server
converts into the renderer's camera convention and applies inside the user's
existing render loop. This keeps integration simple: a method only needs
to write its rendered outputs into \name's buffers and apply client-provided
controls.

\subsection{Clients and applications}

All clients follow the same workflow, opening the shared CUDA buffers, synchronizing
on the CUDA event, displaying synchronized color/depth images, and streaming interaction
commands back to the server. They differ only in how the shared buffers enter
their runtime. For each application, we create a custom plugin using the SplatBus
protocol. We here illustrate the flexibility of SplatBus with three custom clients:
Unity plugin, Blender plugin, OpenGL viewer.

\paragraph{Unity plugin}
The Unity plugin uses native rendering-plugin hooks and
OpenGL--CUDA interop to copy GPU data into Unity render targets, enabling
composition with meshes, and editor interaction. 
When point clouds extracted from Gaussian assets are available, 
they can be imported into the Unity scene and manipulated interactively in the editor view, 
while the corresponding poses are streamed to the server for rendering.
Figure~\ref{fig:unity-screenshot-3dgs} shows a 3DGS scene rendered with
the original 3DGS rasterizer and visualized in Unity through \name. 
The same
mechanism also supports more complex research pipelines such as 4DGS
dynamic-scene pipelines  with neural network components, since the client
consumes only the final GPU color/depth buffers rather than a particular
Gaussian representation.

\paragraph{Blender plugin}
The Blender plugin functions similarly by embedding the (dynamic) scene coloured
point cloud in the viewport, with a render of the scene overlayed on the
viewport. Our plugin instantiates a compositor graph, enabling compositing a
scene onto the 3DGS render. Using the coloured point cloud of the 3DGS scene,
our plugin enables basic ambient lighting and shadow casting over the Blender
scene, as shown in Figure~\ref{fig:blender-screenshot}. In addition, we handle
dynamic 3DGS-based scenes (i.e. 4DGS~\cite{realtime_4dgs}) and synchronize the
Blender timeline with the renderer. This allows users to synchronize mesh
animation with the scene dynamics, and to render composited animations frame by
frame. 

\paragraph{OpenGL viewer}
The OpenGL viewer displays the CUDA buffers as textures without GPU-to-CPU memory copy,
enabling rendering speeds comparable to SIBR~\cite{sibr_framework}.
Due to its minimalist implementation built on ModernGL~\cite{moderngl},
additional features such as profiling and visual debugging of Gaussian splatting
scenes can be implemented with few lines of Python.
Figure~\ref{fig:opengl-screenshot-avatar} shows a trained avatar rendered with 
the original mmlphuman project~\cite{mmlphuman} and visualized in our OpenGL
viewer through \name.

%% file: sec/4_evaluation.tex
\section{Evaluation}

We evaluate SplatBus's renderer-side overhead, end-to-end transfer, and
throughput and fidelity relative to SIBR and two native Unity implementations.
Measurements isolate frame handoff rather than splatting because our focus is
renderer generality and fidelity, not faster rasterization.

\subsection{Experimental setup}

Experiments run on one machine with an NVIDIA RTX 4090 (24\,GB), Intel Core
i9-13900, 32\,GB RAM, and Ubuntu 22.04. We use the original 3DGS CUDA rasteriser
with PyTorch~2.1.2 and CUDA~12.1, and three Mip-NeRF~360 scenes trained for 30k iterations:
\emph{room} (1.6M Gaussians), \emph{garden} (5.8M), and \emph{bicycle} (6.1M).
Results average 10{,}000 frames after 200 warm-up frames while replaying the
dataset cameras, with VSync disabled.

\subsection{Handoff overhead vs.\ the original renderer}
\label{sec:overhead}

With no client attached, we measure the GPU work added on top of the rasteriser:
copying color and depth into shared IPC buffers and recording the completion
event. Because the handoff copies only rendered 2D buffers and never touches the
primitives, its cost is \emph{per-pixel and independent of scene complexity}.
Table~\ref{tab:overhead} confirms that time and memory remain constant across
the three scene sizes at each resolution.

\begin{table}[t]
\centering
\setlength{\tabcolsep}{5pt}
\caption{CUDA-IPC handoff overhead averaged over three Mip-NeRF~360 scenes;
overhead is handoff time divided by render time.}
\label{tab:overhead}
\begin{tabular}{lrrrr}
\toprule
 & 720p & 1080p & 1440p & 4K \\
\midrule
Original render, avg. (ms) & 4.3 & 6.0 & 8.5 & 15.4 \\
$+$ Handoff (ms)           & 0.1 & 0.2 & 0.3 & 1.0  \\
Overhead, avg.             & 2.3\% & 3.3\% & 3.5\% & 6.5\% \\
GPU-mem overhead (MB)      & 7.0 & 15.8 & 28.1 & 63.3 \\
\bottomrule
\end{tabular}
\end{table}

\subsection{Transfer comparison}
\label{sec:transfer}

We compare CUDA-IPC with two alternatives under the same architecture.
\emph{CPU shared memory} copies the frame device-to-host into a RAM buffer, then
host-to-device at the client. \emph{Encoded stream} uses NVENC for lossy color
and 16-bit lossless or 8-bit lossy depth, TCP transfer, and NVDEC. Table~\ref{tab:transfer}
reports 1080p latency on \emph{room} using the same OpenGL client.

The render kernel is identical across modes. CUDA-IPC shares a handle once and
avoids transfer copies, giving the lowest latency and highest throughput. CPU
shared memory incurs two PCIe copies, including a 3.8\,ms device-to-host copy on
the server, while encoded streaming is dominated by the NVENC/NVDEC pipeline.

\begin{table}[t]
\centering
\small
\setlength{\tabcolsep}{3pt}
\caption{End-to-end latency (ms) and throughput (fps). ``Enc.''\ denotes NVENC.
Stage timings use per-frame device synchronization; displayed FPS does not.}
\label{tab:transfer}
\resizebox{\columnwidth}{!}{%
\begin{tabular}{lrrrr}
\toprule
Stage (ms)  & CUDA-IPC & CPU-shm & Enc.\ loss.\ & Enc.\ lossy \\
\midrule
Render               & 5.2  & 5.1  & 6.6  & 5.1  \\
Handoff (server)     & 0.4  & 3.8  & 0.2  & 0.4  \\
Transfer $+$ recv    & 0.4  & 2.9  & 45.8 & 27.4 \\
Viewer copy $+$ draw & 1.9  & 2.1  & 0.7  & 1.2  \\
\midrule
\textbf{Pipeline total} & \textbf{7.9} & \textbf{13.9} & \textbf{53.3} & \textbf{34.1} \\
Displayed FPS        & 169.4 & 118.4 & 52.6 & 92.2 \\
\bottomrule
\end{tabular}%
}
\end{table}

\begin{table*}[t]
\centering
\caption{Displayed throughput (fps) of SIBR, SplatBus clients, and native Unity implementations.}
\label{tab:throughput}
\begin{tabular}{llrrrrrr}
\toprule
 & & & \multicolumn{3}{c}{SplatBus} & \multicolumn{2}{c}{Native Unity} \\
\cmidrule(lr){4-6} \cmidrule(lr){7-8}
Scene & Res. & SIBR
 & \shortstack{OpenGL\\(on-request)}
 & \shortstack{OpenGL\\(free-run)}
 & \shortstack{Unity\\(free-run)}
 & \shortstack{Unity-\\GaussianSplatting}
 & \shortstack{gsplat-\\unity} \\
\midrule
\multirow{4}{*}{room}
 & 720p  & 352.2 & 300.2 & 284.1 & 253.0 & 441.1 & 688.6 \\
 & 1080p & 205.9 & 185.2 & 169.4 & 148.5 & 372.3 & 435.4 \\
 & 1440p & 135.4 & 122.3 & 108.2 & 99.9  & 292.7 & 290.9 \\
 & 4K    & 70.2  & 62.7  & 52.4  & 53.0  & 191.0 & 167.8 \\
\midrule
\multirow{4}{*}{garden}
 & 720p  & 207.5 & 172.6 & 161.3 & 154.1 & 138.5 & 243.9 \\
 & 1080p & 152.8 & 131.5 & 121.6 & 112.2 & 132.9 & 226.9 \\
 & 1440p & 113.2 & 98.2  & 88.1  & 82.4  & 128.2 & 196.1 \\
 & 4K    & 66.9  & 58.0  & 48.5  & 47.7  & 115.3 & 138.1 \\
\midrule
\multirow{4}{*}{bicycle}
 & 720p  & 180.4 & 153.4 & 141.4 & 135.4 & 132.7 & 242.4 \\
 & 1080p & 132.6 & 116.1 & 106.7 & 97.4  & 127.9 & 223.5 \\
 & 1440p & 95.1  & 84.2  & 75.0  & 70.4  & 122.9 & 195.7 \\
 & 4K    & 54.3  & 47.9  & 39.9  & 40.0  & 112.9 & 143.3 \\
\bottomrule
\end{tabular}
\end{table*}

\subsection{Comparison with native viewers}
\label{sec:compare}

We compare the decoupled pipeline against three native rasteriser reimplementations, 
INRIA's standalone SIBR viewer~\cite{sibr_framework},
UnityGaussianSplatting~\cite{pranckevicius2023unitygs}, and gsplat-unity~\cite{gsplatunity}. 
These avoid cross-process handoff but only support static 3DGS.

SplatBus runs two render modes. In \emph{free-run} the server renders
continuously and the client displays the latest published frame. 
This is the mode used by the Unity and Blender clients, whose engine loop must not block. 
In \emph{on-request} the client requests a frame and the server signals when it is ready, matching
SIBR's control flow. 
We use this mode for the OpenGL client to compare with SIBR fairly. 
We set "VeryHigh FP32" for UnityGaussianSplatting, and "Uncompressed" for gsplat-unity.

Tables~\ref{tab:throughput} and~\ref{tab:quality} report throughput and fidelity.
The native viewers rasterize in-process and avoid the cross-process handoff,
so they are generally faster than the decoupled SplatBus clients.
Which native viewer is fastest depends on scene size and resolution.
The Unity reimplementations draw splats as quads through the GPU's
hardware rasteriser, so their per-pixel cost is low, but they re-sort
all splats every frame, a cost that scales with splat count.
SIBR's tile-based rasteriser runs in software (CUDA), which only pays for
visible splats, but its per-pixel cost grows with resolution.
So SIBR leads on large scenes at low resolution, and the Unity
reimplementations lead on small scenes and at 4K.
On fidelity, SplatBus displays the unmodified original rasteriser and leads
PSNR/SSIM/LPIPS on all three scenes.
UnityGaussianSplatting stays within 0.2--0.4\,dB of the original.
gsplat-unity handles small splats more coarsely, which makes it faster
but costs 3--4.5\,dB on the two outdoor scenes.
SplatBus thus trades a modest, bounded overhead for serving an unmodified
research renderer to any client.

\begin{table}[t]
\centering
\setlength{\tabcolsep}{5pt}
\caption{Fidelity on Mip-NeRF 360 test views; SplatBus reproduces the original rasterizer.}
\label{tab:quality}
\begin{tabular}{llccc}
\toprule
Scene & System & PSNR\,$\uparrow$ & SSIM\,$\uparrow$ & LPIPS\,$\downarrow$ \\
\midrule
\multirow{3}{*}{room}
 & \textbf{SplatBus} & \textbf{31.25} & \textbf{0.914} & \textbf{0.221} \\
 & UnityGaussianSplatting & 30.82 & 0.885 & 0.221 \\
 & gsplat-unity          & 30.29 & 0.874 & 0.228 \\
\midrule
\multirow{3}{*}{garden}
 & \textbf{SplatBus} & \textbf{27.04} & \textbf{0.857} & \textbf{0.110} \\
 & UnityGaussianSplatting & 26.87 & 0.852 & 0.113 \\
 & gsplat-unity          & 22.50 & 0.737 & 0.188 \\
\midrule
\multirow{3}{*}{bicycle}
 & \textbf{SplatBus} & \textbf{25.17} & \textbf{0.763} & \textbf{0.212} \\
 & UnityGaussianSplatting & 25.00 & 0.755 & 0.215 \\
 & gsplat-unity          & 21.75 & 0.668 & 0.263 \\
\bottomrule
\end{tabular}
\end{table}

%% file: sec/5_conclusion.tex
\section{Conclusion}

\name decouples Gaussian-based research renderers from interactive viewers by
sharing rendered color and depth buffers through CUDA IPC. This preserves the
original rasterizer output while enabling low-latency OpenGL, Unity, Blender, and other
clients without rewriting the renderer. The result is a practical path for using
new Gaussian variants, such as dynamic scenes, and avatar pipelines inside external
interactive environments. Future work will extend interaction support, rendering
effects, and the client rendering API.